\documentclass[fleqn,usenatbib]{mnras}

\usepackage{newtxtext,newtxmath}

\usepackage[T1]{fontenc}
\usepackage{ae,aecompl}
\usepackage{multicol}

\usepackage{graphicx}	
\usepackage{amsmath}	
\usepackage{amssymb}	

\title[Coeval MPs in NGC 2121]
{Chromosome maps of young LMC clusters: An additional case of coeval multiple populations}

\author[S. Saracino et al.]{S. Saracino$^{1}$\thanks{E-mail: s.saracino@ljmu.ac.uk},
S. Martocchia$^{1,2}$,
N. Bastian$^{1}$,
V. Kozhurina-Platais$^{3}$,
W. Chantereau$^{1}$,
\newauthor
M. Salaris$^{1}$,
I. Cabrera-Ziri$^{4}$\thanks{Hubble Fellow},
E. Dalessandro$^{5}$,
N. Kacharov$^{6}$,
C. Lardo$^{7}$,
S. S. Larsen$^{8}$,
\newauthor
I. Platais$^{10}$
\\
$^{1}$Astrophysics Research Institute, Liverpool John Moores University, 146 Brownlow Hill, Liverpool L3 5RF, UK\\
$^{2}$European Southern Observatory, Karl-Schwarzschild-Stra\ss e 2, D-85748 Garching bei M\"unchen, Germany\\
$^{3}$Space Telescope Science Institute, 3700 San Martin Drive, Baltimore, MD 21218, USA\\
$^{4}$Harvard-Smithsonian Center for Astrophysics, 60 Garden Street, Cambridge, MA 02138, USA\\
$^{5}$INAF-Osservatorio di Astrofisica \& Scienza dello Spazio, via Gobetti 93/3, I-40129, Bologna, Italy\\
$^{6}$Max-Planck-Institut f\"ur Astronomie, K\"onigstuhl 17, D-69117 Heidelberg, Germany\\
$^{7}$Laboratoire d'astrophysique, \' Ecole Polytechnique F\' ed\' erale de Lausanne (EPFL), Observatoire, 1290, Versoix, Switzerland\\
$^{8}$Department of Astrophysics/IMAPP, Radboud University, P.O. Box 9010, 6500 GL Nijmegen, The Netherlands\\
$^{9}$Dipartimento di Fisica \& Astronomia, Universit\` a degli Studi di Bologna, via Gobetti 93/2, I-40129, Bologna, Italy\\
$^{10}$Department of Physics and Astronomy, Johns Hopkins University, 3400 North Charles Street, Baltimore, MD 21218, USA\
}

\date{Accepted XXX. Received YYY; in original form ZZZ}

\pubyear{2019}

\begin{document}
\label{firstpage}
\pagerange{\pageref{firstpage}--\pageref{lastpage}}
\maketitle

\hypersetup{colorlinks=true}

\begin{abstract}
Recent studies have revealed that the Multiple Populations (MPs) phenomenon does not occur only in ancient and massive Galactic globular clusters (GCs), but it is also observed in external galaxies, where GCs sample a wide age range with respect to the Milky Way. However, for a long time, it was unclear whether we were looking at the same phenomenon in different environments or not. The first evidence that the MPs phenomenon is the same regardless of cluster age and host galaxy came out recently, when an intermediate-age cluster from the Small Magellanic Cloud, Lindsay 1, and a Galactic GC have been directly compared. By complementing those data with new images from the Hubble Space Telescope (HST), we extend the comparison to two clusters of different ages: NGC 2121 ($\sim$2.5Gyr) and NGC 1783 ($\sim$1.5Gyr), from the Large Magellanic Cloud. We find a clear correlation between the RGB width in the pseudo-colour $C_{F275W,F343N,F438W}$ and the age of the cluster itself, with the older cluster having larger $\sigma(C_{F275W,F343N,F438W})^{RGB}$ and vice-versa. Unfortunately, the $\sigma$ values cannot be directly linked to the N-abundance variations within the clusters before properly taking account the effect of the first dredge-up. Such HST data also allow us to explore whether multiple star-formation episodes occurred within NGC 2121. The two populations are indistinguishable, with an age difference of only 6$\pm$12 Myr and an initial Helium spread of 0.02 or lower. This confirms our previous results, putting serious constraints on any model proposed to explain the origin of the chemical anomalies in GCs.
\end{abstract}

\begin{keywords}
galaxies: individual: LMC, SMC -- star clusters: individual: Lindsay 1, NGC 2121, NGC 1783 -- technique: photometry
\end{keywords}


\section{Introduction}
Globular clusters (GCs) are no longer considered the best example of Single Stellar Populations (SSP), since much observational evidence has shown that they host star-to-star abundance variations in terms of light elements (e.g. C, N, O, Na).
The Multiple Populations (MPs) phenomenon has been extensively studied in the last years, coming to the conclusion that it is not a peculiarity of Milky Way (MW) GCs but it can be also found in star clusters belonging to external galaxies: Magellanic Clouds (MCs \citealt{mucciarelli2009}; \citealt{milone2009}; \citealt{dalessandro2016}; \citealt{niederhofer2017a}; \citealt{gilligan2019}), Fornax dwarf galaxy \citep{larsen2014} and Sagittarius dwarf galaxies (e.g. M54, \citealt{carretta2010}; \citealt{sills2019}). This finding immediately ruled out the host environment as one of the key elements to understand the phenomenon. Together with mass (\citealt{carretta2010}; \citealt{bragaglia2012}; \citealt{schiavon2013}; \citealt{milone2017}), age plays a crucial role in defining the (observable) onset and properties of chemical anomalies. In the recent years, in fact, MPs have been detected in almost all ancient clusters surveyed \citep{milone2015,milone2017}, as well as in intermediate-age clusters (\citealt{hollyhead2017},\citealt{niederhofer2017a,niederhofer2017b},\citealt{martocchia2018a,martocchia2018b,martocchia2019},\citealt{milone2019}), disappearing for clusters younger that $\sim$2 Gyr (\citealt{martocchia2017}).

The mechanism that triggers the presence of MPs in GCs is still unknown and under debate but its understanding may have important implications for the assembly history of galaxies hosting GC systems (e.g. \citealt{bastianlardo2018}). Several scenarios have been proposed over the years but we still lack a theory able to explain all the recent observational findings. Indeed, depending on the source of the chemical enrichment within the cluster, many theories proposed so far predict age spreads ranging from a few Myr (massive and super-massive stars, e.g. \citealt{decressin2007}; \citealt{denissenkov2014}, \citealt{gieles2018}) up to 30-200 Myr (asymptotic giant branch (AGB) stars, e.g. \citealt{dercole2008}; \citealt{conroy2011}).
An efficient way to rule out one or the other scenarios would be to estimate the age difference among the sub-populations in a cluster. A first attempt in this direction has been done by \citet{marino2012} and \citet[][]{nardiello2015}, where the authors found only upper limits of $\sim$200 Myr, since they focused on ancient GCs. \citet[][]{martocchia2018b} repeated the experiment for the $\sim$2 Gyr cluster NGC 1978 in the Large Magellanic Cloud (LMC), finding an age difference of 1 $\pm$ 20 Myr between the two populations: such a result appears to impose a stringent constraint on the onset of MPs.

This work is the second in a series aimed at investigating the MPs phenomenon in LMC/SMC clusters through specific diagnostics (e.g. the so-called ``chromosome map'' by \citealt{milone2017}), which have been designed for MW GCs but can now be applied to explore different regimes. 
In \citet{saracino2019a} we took advantage of new ultraviolet (UV) data acquired under our ongoing Hubble Space Telescope (HST) survey of LMC/SMC massive clusters, to make the first direct comparison between the SMC intermediate-age cluster Lindsay 1 and the Galactic GC NGC 288, by demonstrating that the MPs formation mechanism should be the same for different galaxies, as well as at different GCs ages.
Here we complement those data with that of two young clusters, namely NGC 2121 and NGC 1783 from the LMC, to analyze in detail how the MPs phenomenon changes with the cluster age. Then we focus our attention on NGC 2121, in order to determine the possible Helium (He) enrichment within the cluster, as well as the possible age difference between its sub-populations.

This paper is organised as follows: in Section \ref{sec:obs} the observational database, the photometric reduction and the analysis are presented. In Section \ref{sec:cmap} an alternative diagnostic for MPs is used to compare Lindsay 1, NGC 2121 and NGC 1783. In Section \ref{sec:helium} we discuss the presence of He-abundance variations in NGC 2121. The analysis of its SGB is then presented in Section \ref{sgb}. We finally summarize the most relevant results and draw our conclusions in Section \ref{sec:concl}.

\section{Observations and data analysis}\label{sec:obs}
\subsection{Data and photometry}
The observations of the three clusters analysed in this work, namely Lindsay 1, NGC 2121 and NGC 1783, are from HST and they consist in both proprietary and archival data from our group, covering a wide wavelength range. 
In particular the F336W, F343N and F438W filters come from the HST survey of LMC/SMC star clusters performed in the last years using the WFC3/UVIS camera (proposals GO-14069 and GO-15062; PI: N. Bastian). These observations have been recently complemented with new data in the UV filter F275W, under the proposal GO-15630 (PI: N. Bastian). 
For NGC 2121, even the F814W images have been collected under this ongoing HST program.
For the other two clusters, we took advantage of archival data from the Advanced Camera for Surveys (ACS) in the F814W filter. In particular, the data for Lindsay 1 come from the program GO-10396 (PI. J. Gallagher), while the data for NGC 1783 have been acquired under the program GO-10595 (PI. P. Goudfrooij). The main properties of the WFC3/UVIS and ACS/WFC images used in the paper, in terms of exposure times in the different filters, are summarized in Table \ref{tab:data}. 
We also point out that the data for Lindsay 1 have not been reduced for this paper, since we used the photometric catalog recently published by \citet{saracino2019a}.

\begin{table*}
\caption{Main properties of the archival and proprietary HST images for Lindsay 1, NGC 2121 and NGC 1783 used in the paper.}
\begin{tabular}{c c c c c l l}
\hline \hline
Cluster & Instrument &    Filter  & Date & N$\times$Exposure Time & Proposal & PI \\
\hline
Lindsay\,1 & UVIS/WFC3 &   F275W  & 2019 & 1500s$+$1501s$+$2$\times$1523s$+$2$\times$1525s  & 15630 & N.\,Bastian \\  
           & UVIS/WFC3 &   F336W  & 2011 & 500s$+$2$\times$1200s  & 14069 & N.\,Bastian \\  
           & UVIS/WFC3 &   F343N  & 2016 & 500s$+$800s$+$1650s$+$1850s       & 14069 & N.\,Bastian \\   
           & UVIS/WFC3 &   F438W  & 2016 &  120s$+$2$\times$460s           & 14069 & N.\,Bastian \\   
           & WFC/ACS   &   F814W  & 2006 &   10s$+$4$\times$474s    & 10396 & J. Gallagher \\
\hline
NGC\,2121 & UVIS/WFC3 &   F275W  & 2019 & 1501s$+$1511s$+$1512s$+$1519s$+$1521s$+$4$\times$1523s$+$1525s+2$\times$1529s & 15630 & N.\,Bastian \\
          & UVIS/WFC3 &   F336W  & 2011 & 270s$+$2$\times$715s  & 15062 & N.\,Bastian \\  
          & UVIS/WFC3 &   F343N  & 2016 & 450s$+$2$\times$1250s$+$1650s       & 15062 & N.\,Bastian \\   
          & UVIS/WFC3 &   F438W  & 2016 &  120s$+$2$\times$550s           & 15062 & N.\,Bastian \\   
          & UVIS/WFC3  &   F814W  & 2019 & 3$\times$200s$+$3$\times$350s$+$700s  & 15630 & N.\,Bastian \\
\hline
NGC\,1783 & UVIS/WFC3 &   F275W  & 2019 & 2$\times$1500s$+$4$\times$1512s & 15630 & N.\,Bastian \\
          & UVIS/WFC3 &   F336W  & 2011 & 2$\times$1190s$+$1200s  & 12257 & L.\,Girardi \\  
          & UVIS/WFC3 &   F343N  & 2016 & 450s$+$845s$+$1650s     & 14069 & N.\,Bastian \\   
          & WFC/ACS   &   F435W  & 2006 &  90s$+$2$\times$340s    & 10595 & P.\,Goudfrooij \\
          & WFC/ACS   &   F814W  & 2006 &   8s$+$2$\times$340s    & 10595 & P.\,Goudfrooij \\
\hline \hline
\end{tabular}
\label{tab:data}
\end{table*}

The photometric analysis has been performed with \texttt{DAOPHOT IV} \citep{stetson87} on images processed, flat-fielded, bias subtracted, and corrected for Charge Transfer Efficiency losses by standard HST pipelines ($\_{flc}$ images). As a first step, few hundreds of stars have been selected in each image and chip in order to model the point spread function (PSF), by considering a 10-pixel aperture. The PSF models were chosen on the basis of a $\chi^2$ statistic and, in average, the best-fit has been provided by a Moffat function \citep{moffat69}. These models were finally applied to all the sources detected at more than $3\sigma$ from the background level in each image. Then, we built a master catalog with stars detected in at least half the available images per filter. In some cases a less restrictive criterion has been adopted in order to cover the gap between the two chips. At the corresponding positions of these stars, the photometric fit was forced in all the other frames by using {\texttt{DAOPHOT IV/ALLFRAME} \citep{stetson94}}. Finally, the magnitude and photometric error of each star has been estimated as the weighted mean and standard deviation of those measured in multiple images.

Instrumental magnitudes have been then reported to the VEGAMAG system by using the zero-points values quoted both in the WFC3 and ACS websites (at the aperture size), as well as appropriate aperture corrections at a radius of 10 pixels from the stellar centers.
Instrumental positions have been reported to the absolute coordinate system (RA, Dec) by using the stars in common with the Gaia Data Release 2 (DR2, \citealp{gaia2016,gaia2018}) and by means of the cross-correlation software \texttt{CataXcorr} \citep{montegriffo1995}. This software has been also used to combine the final WFC3 and ACS catalogues of the clusters.

\subsection{Background decontamination \& Differential Reddening}
From previous studies of LMC/SMC clusters (e.g. \citealt{dalessandro2019}), we have learned that the contamination from field stars could be relevant in these regions, contributing in some way to alter the scientific results. It appears to be not the case for Lindsay 1, where the contribution is relatively low \citep{saracino2019a}. For NGC 2121 and NGC 1783, we instead adopted a statistical approach, which allows to estimate the fraction of star interlopers along the main evolutionary sequences in a colour-magnitude diagram (CMD). In particular we applied the method described by Cabrera-Ziri et al. submitted, which has already been successfully tested on the SMC cluster NGC 419. This method is similar to the one described by \citet{niederhofer2017a} and commonly used in the previous papers of this series.

Since no parallel fields are available in the HST archive for NGC 2121 and NGC 1783, we then defined as cluster region, the first 40'' (45'' for NGC 1783) from the cluster center\footnote{The cluster centers of NGC 2121 and NGC 1783 have been determined by following the iterative procedure described in \citet[][see also \citealt{ferraro2003}; \citealt{lanzoni2007}]{montegriffo1995}, which consists in averaging the absolute positions of properly selected stars at different distances from a first-guess center. (RA, Dec) = (87.059036\textdegree, Dec=-71.479805\textdegree) and (RA, Dec) = (74.7880\textdegree, Dec=-65.8965\textdegree) are the results for NGC 2121 and NGC 1783, respectively.}, while as control field region, the one made by all the stars located at a distance greater than 75'' (80'' for NGC 1783) from the center. Since the area ratio between cluster and control field is $\approx$ 0.6, for every star in the control region, we flagged $\approx$ 0.6 stars in the cluster region as likely members of the field according to its distance to the control field stars and the relative uncertainties. We repeated the same procedure 1000 times and at the end, we cleaned the cluster region by removing all the stars that have been flagged as field members $>$ 75\% of the times. For a detailed description of the method we refer to Cabrera-Ziri et al., submitted.

\begin{figure*}
    \centering
	\includegraphics[width=\textwidth]{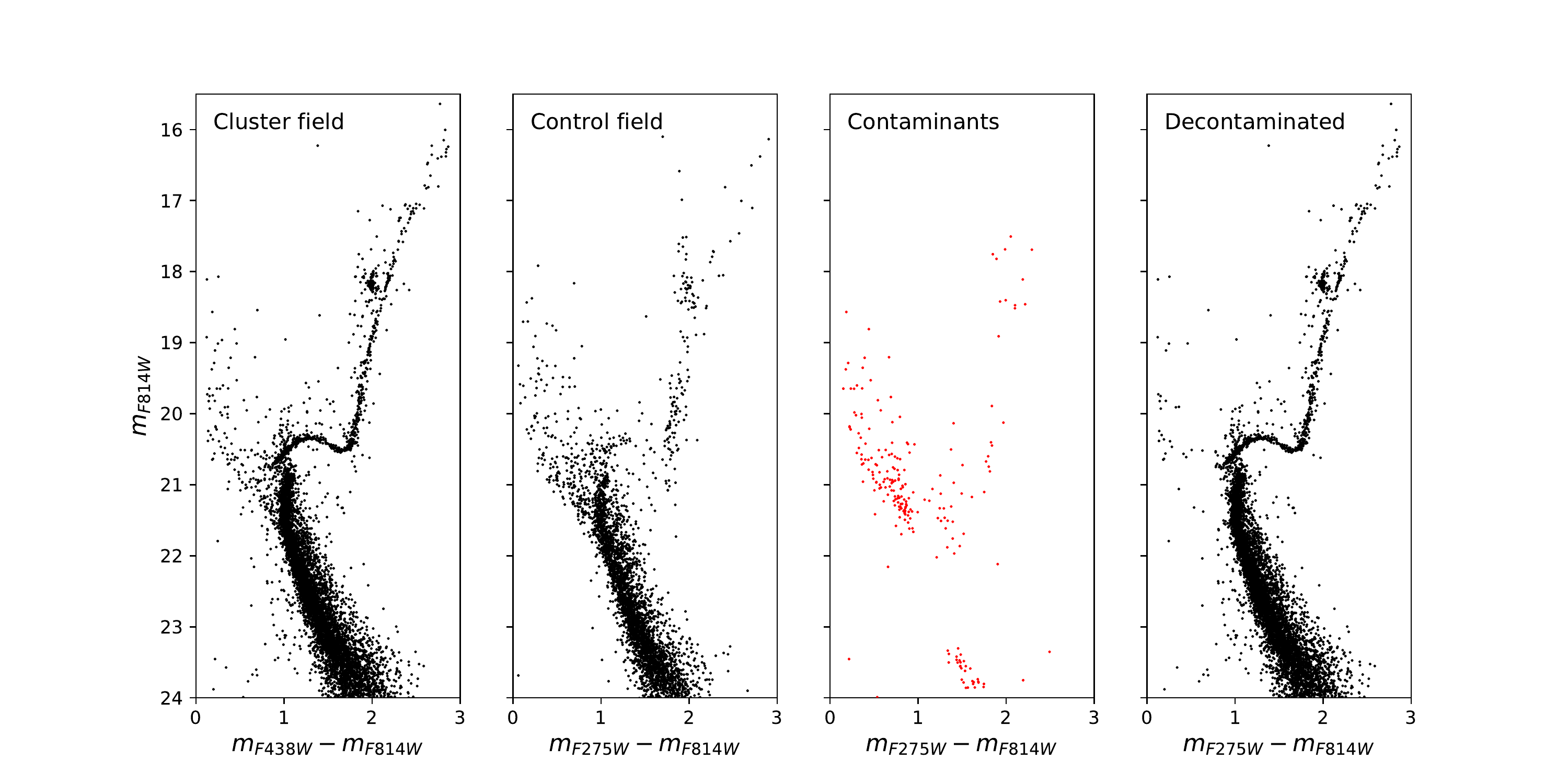}
    \caption{The first two panels represent the $m_{F814W}$, $m_{F438W}-m_{F814W}$) CMDs of the cluster region and the control field of NGC 2121 respectively, while the rightmost panel shows how the CMD of NGC 2121 looks like once the background decontamination is applied. The stars flagged as likely field members > 75\% of the times are shown in red in the Figure and then removed from the final sample.}
    \label{fig:cmd_deco}
\end{figure*}

The background subtraction procedure for NGC 2121 is presented as an example in Figure \ref{fig:cmd_deco}. By comparing the cluster field with the decontaminated one (leftmost and rightmost panels, respectively), most of the stars lying on the blue of the MS have been rejected, as well as some spurious stars lying on the SGB and RGB sequences, thus producing a much cleaner and more defined diagram. The output in the case of NGC 1783 is almost the same. The CMDs used in the rest of the paper refer to the innermost and background-subtracted portion of the clusters. 

By inspecting the rightmost CMD of Figure \ref{fig:cmd_deco}, it is clear that NGC 2121 is only slightly affected by differential reddening across the FOV exploited in this study, since all the evolutionary sequences are very well defined. In \citet{saracino2019a}, the authors came to the same conclusions for Lindsay 1 and it appears to be quite common for such LMC/SMC clusters \citep{martocchia2018a,martocchia2019}. Anyway, we estimated the differential reddening effect in NGC 2121 and NGC 1783 in order to be consistent with previous works. To do so, we used the method presented in \citet[][see also \citealt{dalessandro2018}]{saracino2019b}. Briefly, we first created the cluster mean ridge line (MRL) in the ($m_{F814W}$, $m_{F438W}-m_{F814W}$) CMD, then we selected a sample of bona-fide stars in the magnitude range 19.5$<m_{F814W}<$22.5 and we computed the geometrical distance ($\Delta X$) of those stars from the MRL. This reference sample has been then used to assign a $\Delta X$ value to each star in our photometric catalogue, by looking at its 30 closest reference stars. Using the extinction coefficients from \citet{cardelli1989}, we finally transformed $\Delta X$ of each star into the local differential reddening $\delta E(B-V)$. The resulting $\delta E(B-V)$ are very low for both clusters (a mean value of 0.001 and a maximum variation of about 0.01 in the FOV), thus not producing any appreciable difference in their CMDs. As a double check, the technique explained in \citet{milone2012} has been also used, coming to the same conclusions.  

\section{Chromosome Maps}\label{sec:cmap}
One of the main tools adopted in the last years to look onto the MPs phenomenon in GCs is the so-called ``chromosome'' map \citep{milone2017,milone2018a}, a pseudo-color diagram which exploits a combination of F275W, F336W, F438W, and F814W filters to separate populations having different light-element abundances. The power of such a combination comes from two aspects: {\it 1)} the colour combination ($m_{F275W}-m_{F814W}$) on the x-axis is mostly sensitive to temperature variations, and so to He-abundance variations among different stellar populations; {\it 2)} the filter combination ($m_{F275W}-m_{F336W}) - (m_{F336W}-m_{F438W}$) = $C_{F275W,F336W,F438W}$ on the y-axis is mainly a measure of the N-abundance variations within the cluster. The MPs phenomenon of Galactic GCs has been extensively investigated via the chromosome map in the last years thanks to the UV Legacy Survey of Galactic GCs \citep{piotto2015,nardiello2018}. It allowed the identification of, for example, the interesting trends between the light-elements enhancement in a cluster and its total mass. However almost all the GCs in the MW are essentially old (from 10 to 13 Gyr), thus representing one of the main limitations in putting all these findings in a more general framework.
In this respect, the LMC/SMC survey of young and intermediate age clusters allowed us to investigate the phenomenon by looking into a different parameter space: the cluster age. 

In particular, using a slightly different diagnostic, \citet[see also \citealt{niederhofer2017a,niederhofer2017b}]{martocchia2018a,martocchia2019} have found a clear correlation between the N-enhancement and the cluster age, with no evidence of MPs for clusters younger that $\sim$2 Gyr. These studies have also demonstrated the power of the F343N filter of the WFC3/UVIS to infer the N-content of stars in clusters. 
A direct comparison of these two approaches has been recently performed in \citet[][]{saracino2019a}, where we presented the first chromosome map of the extragalactic 7.5 Gyr-old \citep{glatt2008} cluster Lindsay 1, demonstrating that they are both very effective methods for the detection of MPs in clusters.
Here we complement the results achieved for Lindsay 1, with what derived for NGC 2121 and NGC 1783, having very different ages: $\sim$2.5 Gyr \citep{martocchia2019} and $\sim$1.5 Gyr \citep{cabrera2016,mucciarelli2007}, respectively, by using a combined approach which exploits the power of the F343N filter in constructing the chromosome map of the clusters.
This idea is also supported by Figure 3 of \citet[][]{milone2019}, where the authors show that by substituting the F343N filter with the standard F336W, the distinction between the populations becomes even more pronounced.
NGC 2121 has been discovered to host MPs by \citet[][see also \citealt{li2019}]{martocchia2019}, while NGC 1783 does not show any evidence of MPs from both photometric and spectroscopic studies \citep{mucciarelli2007,cabrera2016,zhang2018}.

\begin{figure*}
    \centering
	\includegraphics[width=0.85\textwidth]{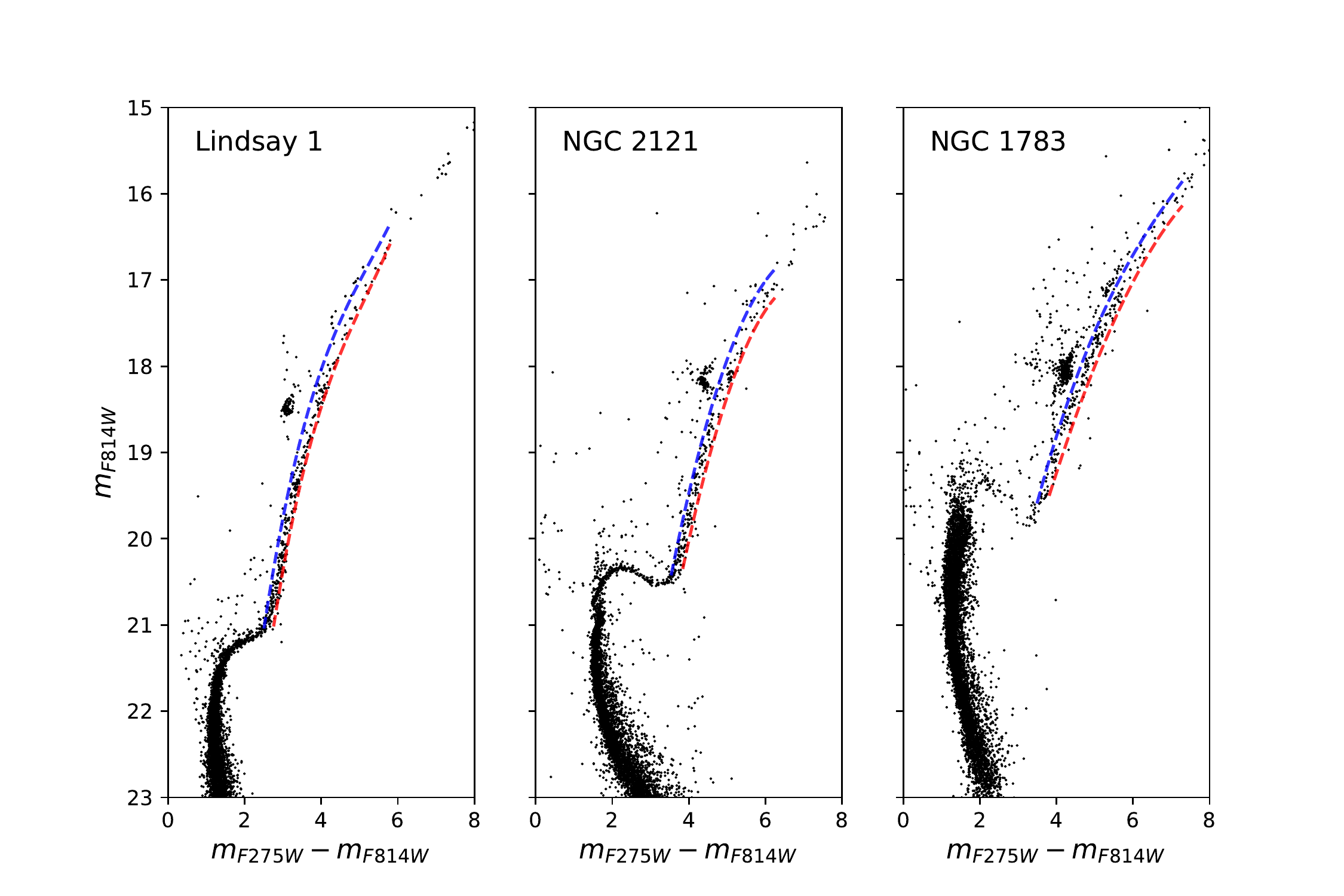}
    \caption{($m_{F814W}$, $m_{F275W}-m_{F814W}$) CMD for all the stars of Lindsay 1 (left panel), NGC 2121 (middle panel) and NGC 1783 (right panel) used in this work. Blue and red dashed lines represent the adopted fiducial lines in the analysis (see Section \ref{sec:cmap} for details).}
    \label{fig:color}
\end{figure*}

To build the chromosome map of the three clusters we have used the procedure outlined in \citet{milone2017}, but adopting the F343N filter instead of the F336W one.
Briefly, we first used the ($m_{F814W}$, $m_{F438W}-m_{F814W}$) CMD to select bona-fide RGB stars. Then we used the ($m_{F814W}$, $m_{F275W}-m_{F814W}$) CMD to define two fiducial lines as the 5th and 95th percentiles of the ($m_{F275W}-m_{F814W}$) distribution of the previously selected RGB stars. We then verticalised the distribution of RGB stars and normalized them to the intrinsic RGB width at 2 mag brighter than the turn-off in the F814W filter, thus creating the $\Delta_{F275W,F814W}$. The ($m_{F814W}$, $m_{F275W}-m_{F814W}$) CMDs of the clusters are presented in Figure \ref{fig:color}, where the fiducial lines are highlighted in red and blue, in all the panels.
We applied the same approach to the pseudo-colour diagram ($m_{F814W}$, $C_{F275W,F343N,F438W}$\footnote{$C_{F275W,F343N,F438W}$ = ($m_{F275W}-m_{F343N}) - (m_{F343N}-m_{F438W}$) as defined in \citet{milone2019}. It is worth to mention that for NGC 1783 we are dealing with F435W instead of F438W.}), in order to derive $\Delta_{F275W,F343N,F438W}$. 

\begin{figure*}
    \centering
	\includegraphics[width=0.85\textwidth]{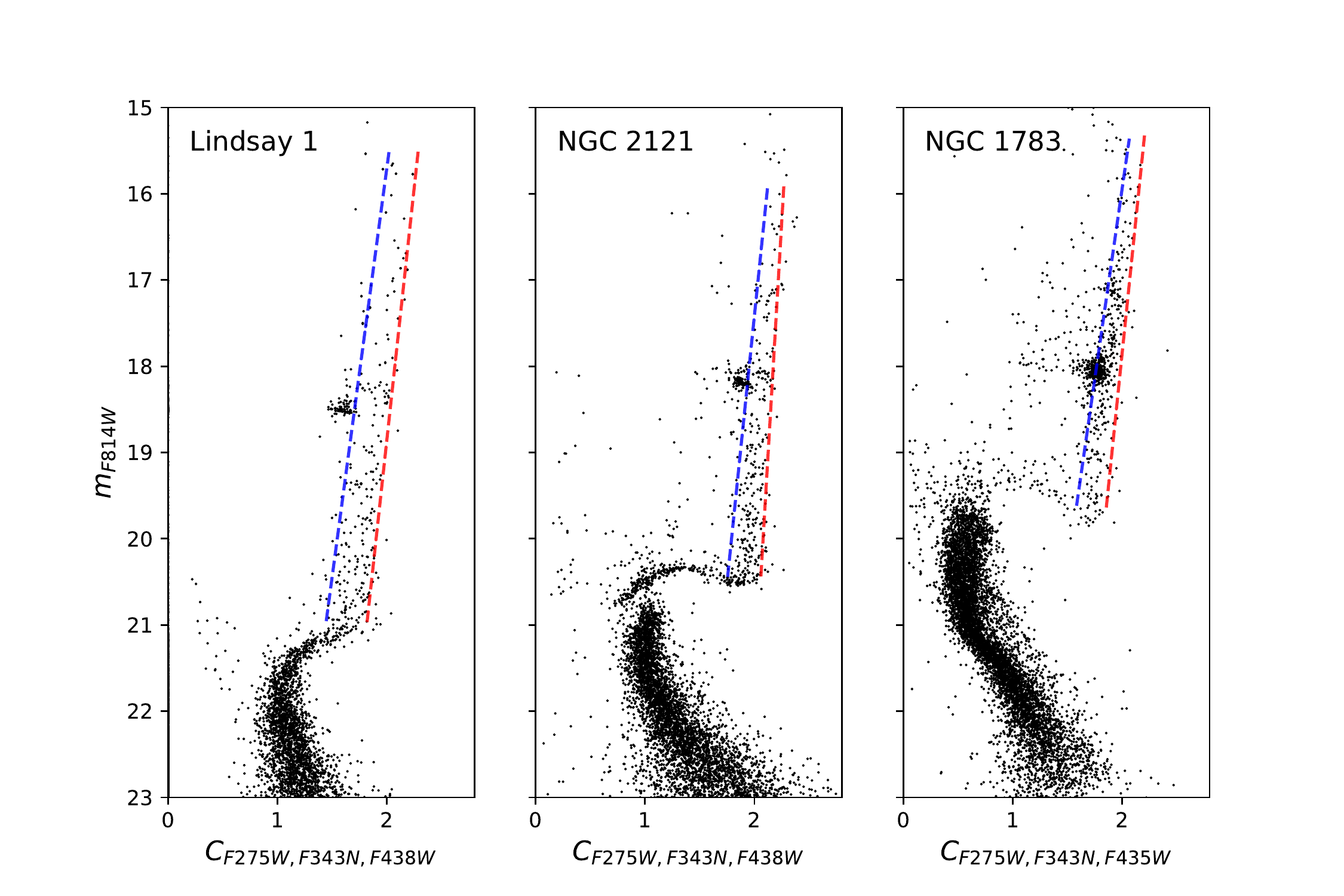}
    \caption{($m_{F814W}$, $C_{F275W,F343N,F435/8W}$) CMD for all the stars of Lindsay 1 (left panel), NGC 2121 (middle panel) and NGC 1783 (right panel) used in this work. Blue and red dashed lines represent the adopted fiducial lines in the analysis.}
    \label{fig:pseudo}
\end{figure*}

\begin{figure*}
    \centering
	\includegraphics[width=\textwidth]{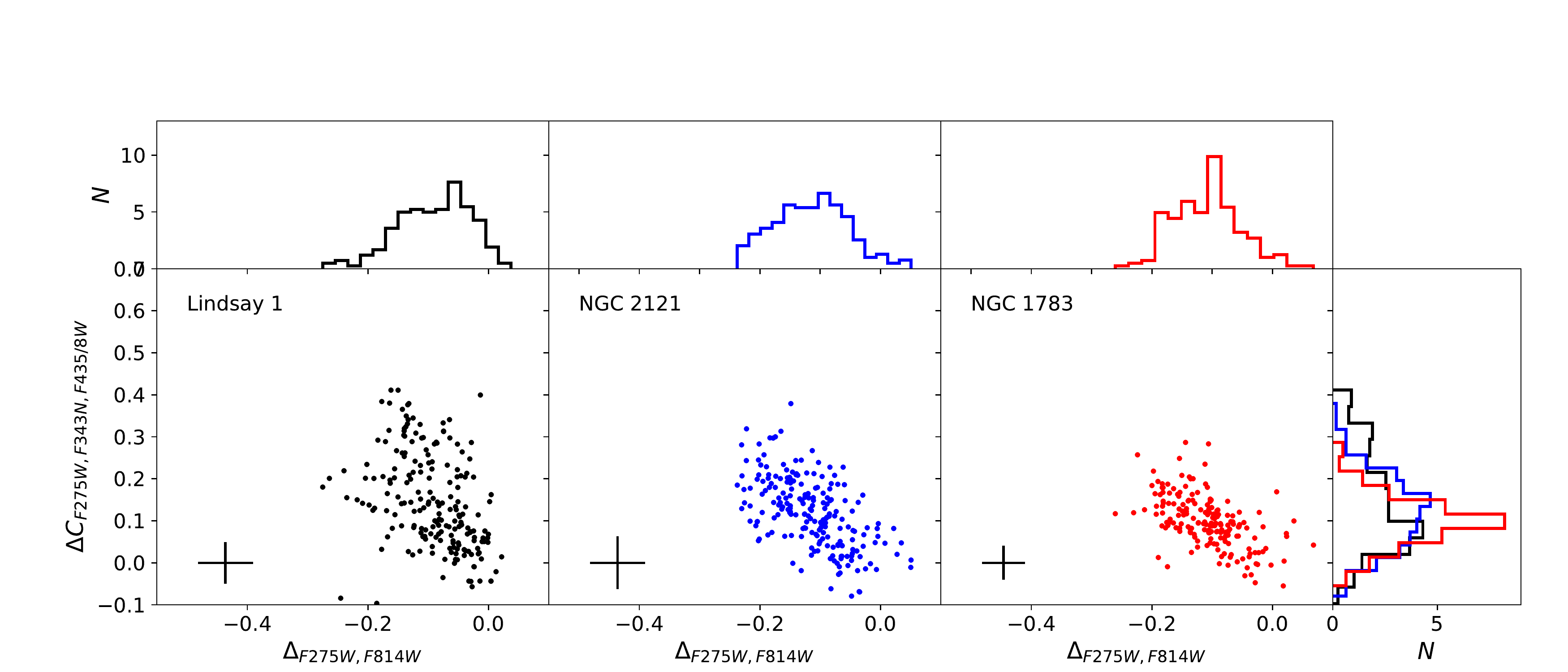}
    \caption{Comparison of the ($\Delta_{F275W,F814W}$, $\Delta_{F275W,F343N,F435/8W}$) chromosome maps for Lindsay 1 (left, black points), NGC 2121 (middle, blue points) and NGC 1783 (right, red points). The histograms of the verticalised colour distribution $\Delta_{F275W,F814W}$ for the three clusters are shown in the top panels, colour-coded as the main ones. In the rightmost panel instead the histograms of the verticalised pseudo-colour distribution $\Delta_{F275W,F343N,F438W}$ for the three clusters are presented one on top of the others. A clear correlation between $\Delta_{F275W,F343N,F438W}$ and the cluster age is visible.}
    \label{fig:chM_delta}
\end{figure*}

The pseudo 
($m_{F814W}$, $C_{F275W,F343N,F438W}$) CMDs of Lindsay 1, NGC 2121 and NGC 1783 are shown in Figure \ref{fig:pseudo}, where the fiducial lines are colour-coded as in Figure \ref{fig:color}.
These values have been used to compute the ($\Delta_{F275W,F814W}$ vs. $\Delta_{F275W,F343N,F438W}$) chromosome map of the three clusters presented in Figure \ref{fig:chM_delta}: Lindsay 1 as black points in the left panel, NGC 2121 as blue points in the middle panel and NGC 1783 as red points in the right panel. The histograms of $\Delta_{F275W,F814W}$ and $\Delta_{F275W,F343N,F435/8W}$ are reported in the ancillary panels, where the same colour-code as for the main panels is used. Figure \ref{fig:chM_delta} shows that Lindsay 1, NGC 2121 and NGC 1783 share exactly the same trend in such a parameter space, with almost the same inclination angle\footnote{The slope of a cluster's chromosome map slightly changes as a function of the metallicity of the cluster itself, with steeper slopes associated to more metal-poor clusters.}. They have different extensions to each other but the observed distribution of stars of each cluster is always wider than what expected from photometric errors alone (shown in the three main panels, bottom-left side). The observed spread clearly indicates the presence of MPs in the clusters, however the separation between the sub-populations looks here very small, contrary to what happens for most of the chromosome maps of MW GCs, thus preventing to distinguish between them.

The comparison of the chromosome map of the three clusters also clearly suggests that the width of the $\Delta_{F275W,F343N,F435/8W}$ distribution strongly correlates with the cluster age, since it decreases going from intermediate-age clusters like Lindsay 1 to young clusters, like NGC 1783. This has been confirmed by computing the $\Delta_{F275W,F343N,F435/8W}$ width for the three clusters: Lindsay 1, $\sigma(C_{F275W,F343N,F438W})^{RGB}_{Lindsay1}$ = 0.12, NGC 2121, $\sigma(C_{F275W,F343N,F438W})^{RGB}_{NGC2121}$ = 0.08, and NGC 1783, $\sigma(C_{F275W,F343N,F435W})^{RGB}_{NGC1783}$ = 0.06. This result is not unexpected since in \citet[][]{martocchia2019} the authors came to the same conclusion by analyzing the $\Delta_{F336W,F438W,F343N}$ and $\Delta_{F343N,F438W,F814W}$ pseudo-colors for a much larger sample of LMC/SMC clusters.

In this regard it is worth to mention the work by \citet[][]{Salaris2020}, where the authors have investigated the effect of the physical process called ``first dredge-up'' on the chemical mixing (in terms of N and other light-elements) of RGB stars in star clusters, concluding that the observed $\sigma(\Delta_{F275W,F343N,F435/8W})^{RGB}$ (as well as that derived from other pseudo-colours) are always a lower limit to the real values and that the discrepancy between real vs. observed widths becomes much more severe for young clusters than old/intermediate age ones. As a direct consequence, all the N abundance spreads derived from the measured $\sigma$ suffer from this effect, so that the results cannot be corrected in a trivial way.
The effect of the first dredge-up is also visible in the pseudo-colour diagrams of Figure \ref{fig:pseudo}, where the RGB width of the clusters decreases going from faint to bright magnitudes. 

In order to quantify such a variation we performed the following test: for each cluster we measured the RGB width at two different magnitude levels, the RGB base and 3 magnitudes above this level. This choice has been made to sample the same evolutionary stage in clusters having very different ages. The comparison between these two values gives an idea of how much the RGB width varies due to the effect of the first dredge up in each cluster. This ratio turns out to be 84.1\% for Lindsay 1, 66.2\% for NGC 2121 and 64.5\% for NGC 1783. As we can see, in agreement with what expected, this effect is much more evident in young clusters than in old ones, thus suggesting that the criterion used to date to normalize the RGB width (at 2 mag above the turn-off of the cluster) for deriving the chromosome maps should be in some way revised as a function of the cluster age.

\section{Helium enrichment in NGC 2121}\label{sec:helium}
In a cluster hosting MPs, stars with slightly different light-element abundances (e.g. C, N, O, Na) are expected to show also He-abundance variations. As already pointed out in the previous section, the distribution of RGB stars in the ($m_{F275W}-m_{F814W}$) colour combination is commonly used as a tracer of such a spread since it strongly depends on the temperature. However other effects (e.g. binaries) can play a similar role in this filter combination, thus affecting in some way the results. A clear example is presented in Figure \ref{sec:cmap} where Lindsay 1, NGC 2121 and NGC 1783 show almost the same width in the x-axis regardless their age and metal content, so that a relative comparison is not straightforward. 

A better way to independently measure the He enrichment within a cluster consists in analyzing the morphology of its red clump (RC) and directly comparing it with appropriate synthetic RC models, as already done for a few LMC/SMC clusters by \citet{chantereau2019}. 
Here we used the same method, applied to the ($m_{F438W}$, $m_{F438W}-m_{F814W}$) CMD, in order to investigate the presence of a He enrichment in NGC 2121 - we refer the reader to \citet{chantereau2019} for more details on the procedure adopted. 

The RC analysis revealed that, in such an age regime, the contribution of a He enrichment ($\Delta$Y$_{ini}$) and of a differential mass-loss on the RGB ($\delta \eta_{R}$) of the cluster cannot be easily disentangled since they almost go in the same direction. This finding, coupled with the photometric errors of the observations ($\sim$0.01 mag both in magnitudes and colours), makes it difficult to draw firm conclusions about the existence of a He enrichment in NGC 2121. 
Indeed, an initial He spread (at fixed mass-loss) or a mass-loss spread (at fixed Y$_{ini}$) can similarly approximate the colour extension and slope of the observed RC.

\begin{figure*}
\begin{multicols}{2}
\includegraphics[width=\linewidth]{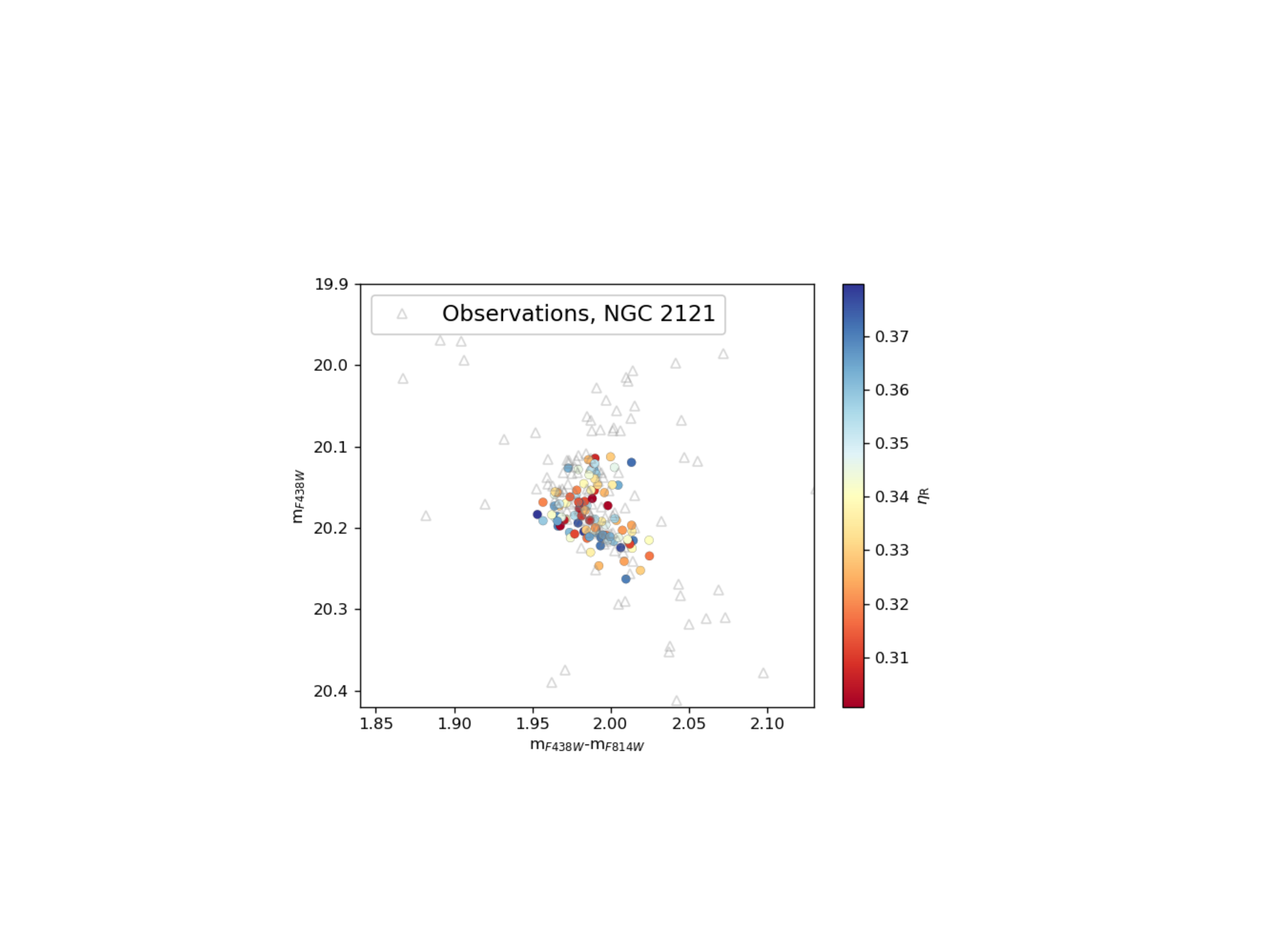}\par 
\includegraphics[width=1.0\linewidth]{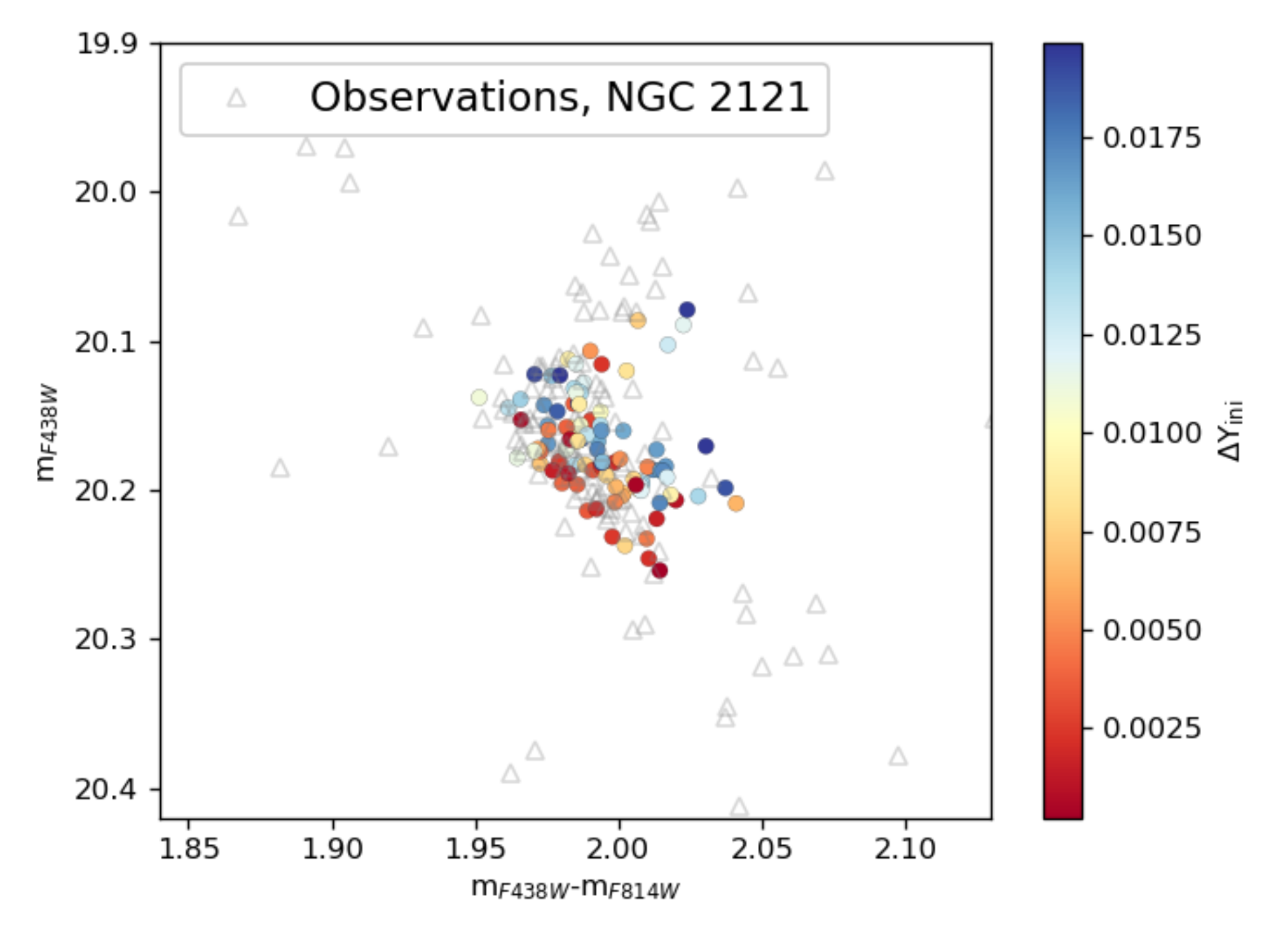}\par 
\end{multicols}
\caption{Zoom into the RC region of NGC 2121 for the ($m_{F438W}$, $m_{F438W}-m_{F814W}$) CMD. The observed stars are shown as grey open triangles in both panels. \textit{Left panel}: Synthetic RC stars with $\Delta$Y$_{ini}$ = 0, colour-coded as a function of $\eta_{R}$ (corresponding to a differential RGB mass-loss $\Delta M_{RGB}$ $\sim$ 0.03-0.04~M$_\odot$). This model only poorly fits the full extension of the RC, thus suggesting that an initial He spread should be present within the cluster. \textit{Right panel}: Synthetic RC stars with $\Delta$Y$_{ini}$ up to 0.02 and a constant RGB mass-loss of $\sim$0.13~M$_\odot$ are presented as filled circles, colour-coded according to the He enhancement. The RC slope and color extension are very well reproduced here. \label{fig:hb_n2121}}
\end{figure*}

We found that a mass-loss parameter in the Reimers formula of about $\eta_R$ = 0.38 (corresponding to a total RGB mass-loss $\sim$0.13~M$_\odot$), is needed to fit the RC position\footnote{We imposed that the reddest edge, in colour, of the synthetic RC overlaps with that of the observed RC.} of NGC 2121 and if the colour extension of the RC is caused by differential mass-loss, then a total RGB mass-loss spread $\Delta M_{RGB}$ of 0.03-0.04~M$_\odot$ (decreasing from $\eta_R$ = 0.38 to 0.30) would be necessary to reproduce its whole extension. 
To support our visual inspection, a statistical comparison between the observed and the modeled RC has been done through a two-dimensional Kolmogorov-Smirnov (2D-KS) test. It turns out that such a range of $\eta_R$ provides the best possible match to our observations.
This is a relatively large amount considering that it roughly corresponds to the 30\% of the total RGB mass-loss in the cluster but there is no way to rule out this scenario. 
It is shown in the left panel of Figure \ref{fig:hb_n2121}, where the observed RC of NGC 2121 in the ($m_{F438W}$, $m_{F438W}-m_{F814W}$) CMD is compared with synthetic RC stars having $\Delta$Y$_{ini}$ = 0 and colour-coded as a function of the mass-loss $\eta_{R}$.

Alternatively, if we make the assumption that the mass-loss along the RGB is essentially constant, which is consistent with what has been found in most stellar clusters studied so far in the Magellanic Clouds \citep{chantereau2019}, then the observations are no longer compatible with models having no initial He spread. This issue can be successfully overcome by invoking a maximum He enrichment of up to 0.020 $\pm$ 0.005 within NGC 2121, for a total RGB mass-loss of $\sim$0.13~M$_\odot$. The value of $\Delta$Y$_{ini}$ suggested for NGC~2121 (M = 0.9$\times 10^5$~M$_\odot$, \citealt{mclaughlin2005}) would follow the relation observed for MW GCs and MCs clusters between $\Delta$Y$_{ini}$ and the mass of the cluster itself \citep{milone2018b,baumgardt2018}.

The right panel of Figure \ref{fig:hb_n2121} shows such a case, where the observed cluster RC well overlaps with synthetic RC stars colour-coded as a function of $\Delta$Y$_{ini}$. Both the slope and the color extension of the RC are quite well reproduced in this scenario.
As before, our choice of $\Delta$Y$_{ini}$ = 0.02 comes from the result of a 2D-KS test: both lower and higher initial He spreads provide significantly worse matches to the data.

From a statistical point of view, the two presented scenarios are indistinguishable so that we are not able to decide which one we should prefer. We note also that the values of $\Delta M_{RGB}$ and $\Delta$Y$_{ini}$ found in this section should be considered as upper limits since a combination of these two effects (both He and mass loss variation) can also be in place in NGC 2121.

\section{Age difference at the SGB level of NGC 2121}\label{sgb}
In this section we focus on the SGB morphology of NGC 2121, in order to investigate the presence of an age difference between the two populations belonging to the cluster, with different N-abundances. To do so we follow the approach adopted by \citet[][]{martocchia2018b} for NGC 1978. This is an interesting aspect to look at, since the results can help to put constraints on the proposed formation mechanisms for the MPs phenomenon.
Lindsay 1 and NGC 1783 have been excluded from the analysis based on their ages: Lindsay 1 is relatively old, so that the morphology of its SGB is not ideal to distinguish stars having different N-abundances, while NGC 1783 is younger than 2 Gyr, the observed age threshold for clusters hosting MPs, so that if an age difference is present, it is expected to be sensibly small.

\subsection{SGB selection}\label{sgb1}
\begin{figure}
    \centering
	\includegraphics[width=0.4\textwidth]{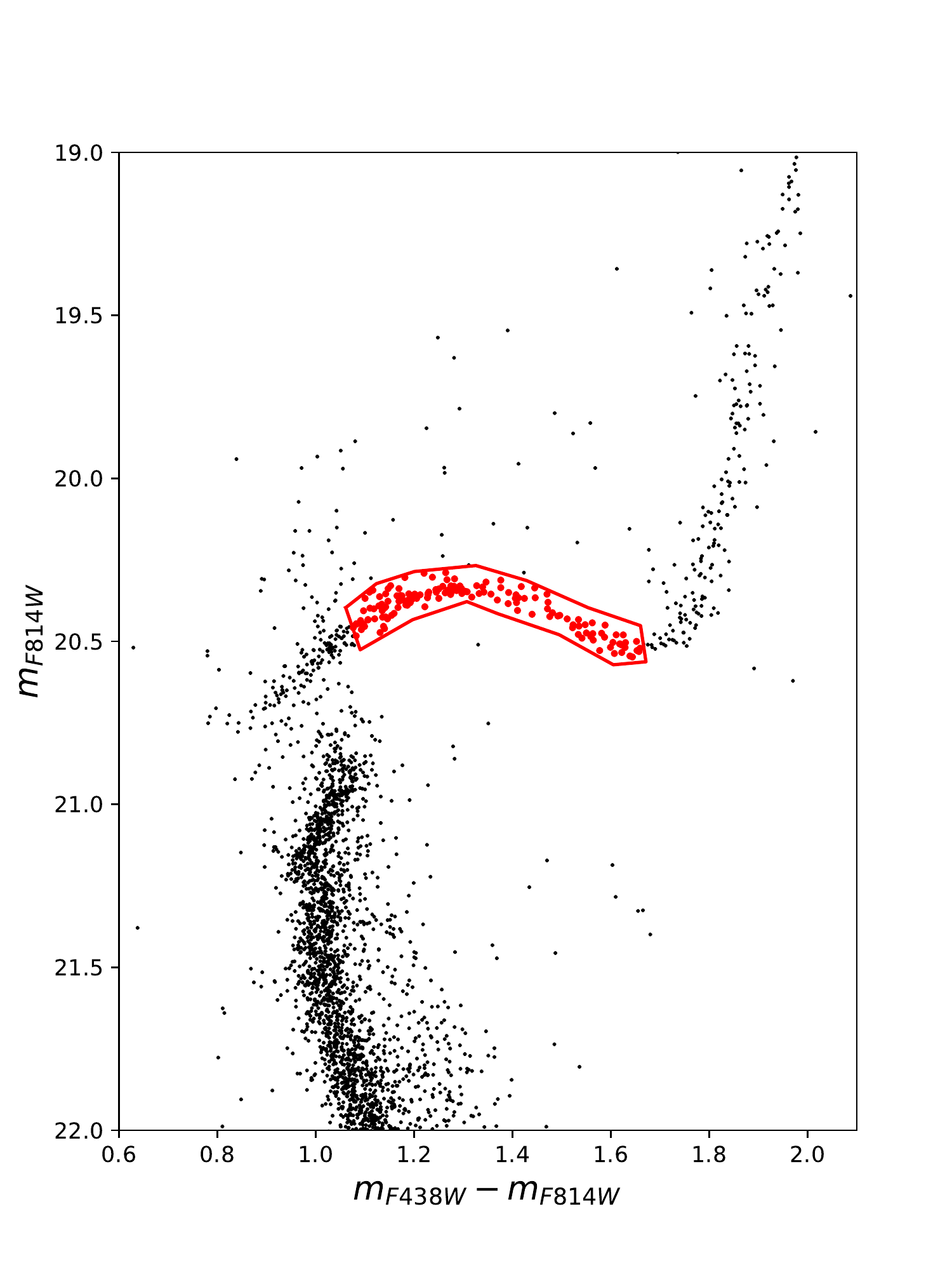}
    \caption{($m_{F814W}$, $m_{F438W}-m_{F814W}$) CMD of NGC 2121. The red box indicates the locus of the initial selection of SGB stars which are marked by red filled circles.}
    \label{fig:SGBsel}
\end{figure}

Briefly, we first used the ($m_{F814W}$, $m_{F438W}-m_{F814W}$) CMD to select the SGB stars in the color range 1.1$<m_{F438W}-m_{F814W}<$1.7, according to the red box in Figure \ref{fig:SGBsel}. Then we used the ($m_{F438W}$, $m_{F343N}-m_{F438W}$) CMD to investigate the possible presence of a bimodal distribution for the previously selected stars. Indeed, in \citet[][]{martocchia2018b}, the authors have demonstrated that two populations having different chemical mixtures (for example N-normal and N-enriched) lie in a slightly different position in the ($m_{F438W}$, $m_{F343N}-m_{F438W}$) colour-magnitude space in the SGB, with the N-enriched population appearing redder than the N-normal one - refer to Figure 6 of \citet[][]{martocchia2018b} for more details.

\begin{figure}
\centering
\includegraphics[width=0.47\textwidth]{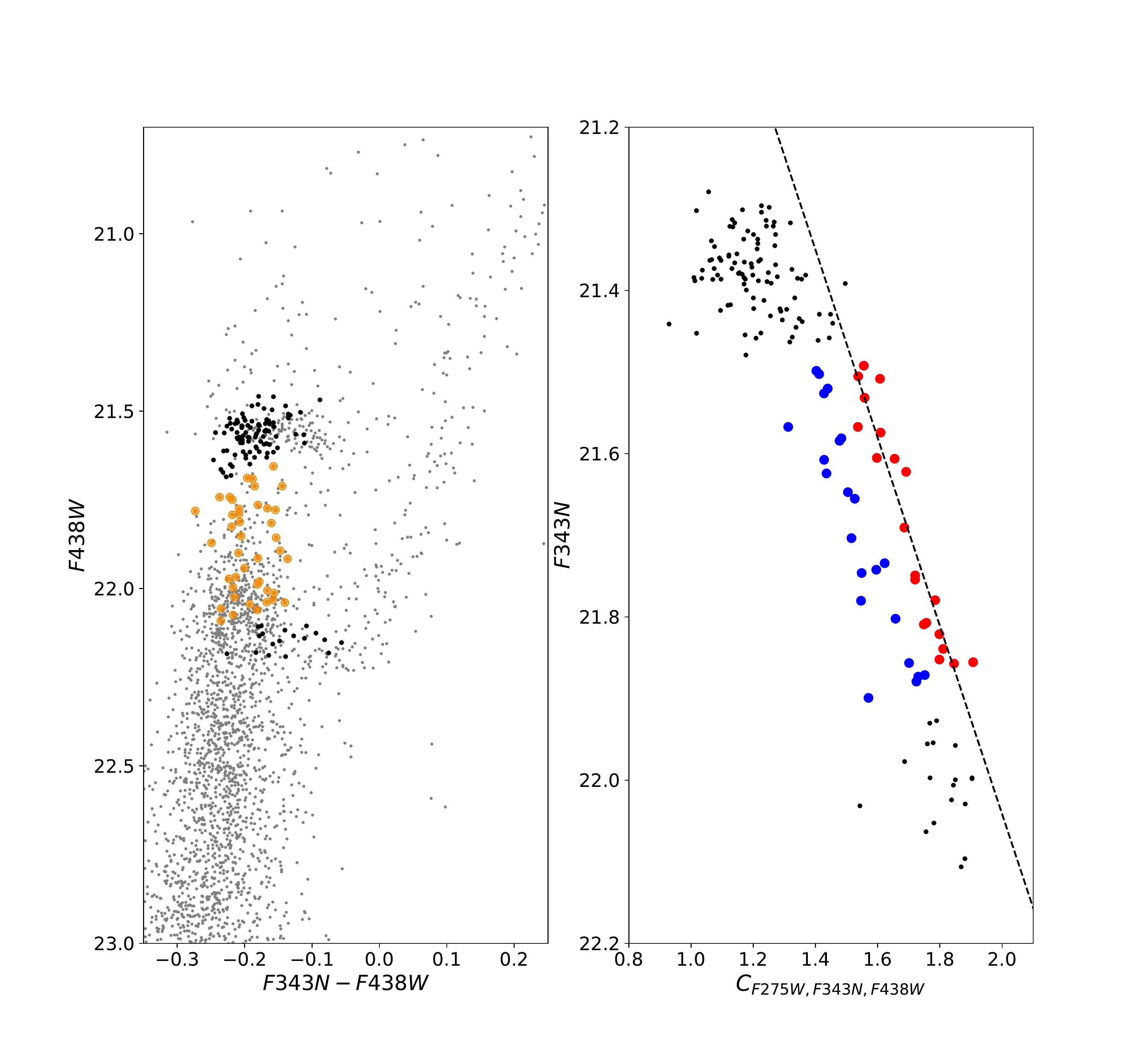}
\caption{{\it Left panel:} The ($m_{F438W}$, $m_{F343N}-m_{F438W}$) CMD of NGC 2121 is shown in grey, with all the previously selected SGB stars overimposed as black circles. Orange filled circles represent the final selected SGB stars. {\it Right panel:} ($m_{F343N}$, $C_{F275W,F343N,F438W}$) pseudo-color CMD of the first selected SGB stars (in grey) of NGC 2121. The final selected SGB stars, shown as larger circles, occupy two well defined sequences in this diagram, one associated to a N-normal population and the other to a N-enriched one. To better visualize the two samples we colored them as red and blue circles, respectively. The black dashed line represent the best-fit (fiducial line) of the red circles in the figure.}
\label{fig:SGBdelta}
\end{figure}

The left panel of Figure \ref{fig:SGBdelta} shows in grey the ($m_{F438W}$, $m_{F343N}-m_{F438W}$) CMD of NGC 2121, while the SGB stars selected in the optical CMD and highlighted in red in Figure \ref{fig:SGBsel} are superimposed here as black circles. As can be seen, in this filter combination the SGB has a peculiar shape, partially overlapping with the main-sequence. We then focused on the almost vertical part of the SGB, in the magnitude range 21.6$<m_{F438W}<$22.1, to make the final selection of SGB stars. In this region the observed sequence shows a hint of bimodality.
The final selected SGB stars are shown as orange filled circles in Figure \ref{fig:SGBdelta}, left panel. We will call hereafter the SGB stars survived to the last selection as final SGB stars.

To properly separate the two populations, we tested once again the power of the pseudo-colour ($C_{F275W,F343N,F438W}$), but this time at the SGB level. This colour combination appears to work relatively well also in this case and the result is shown in the right panel of Figure \ref{fig:SGBdelta}, where the final SGB stars define two distinct sequences. The two populations have been identified and highlighted as red and blue points in the figure. It is worth to mention that, according to the definition of the pseudo-colour diagram, the x-axis is somewhat flipped so that the N-normal population lies on the red side of the diagram, while the N-enriched population on the blue one.
By using the red points in Figure \ref{fig:SGBdelta} we defined a fiducial line for the first population as a linear best-fit, which is also displayed in the figure as a black dashed line. 

\begin{figure}
    \centering
	\includegraphics[width=0.35\textwidth]{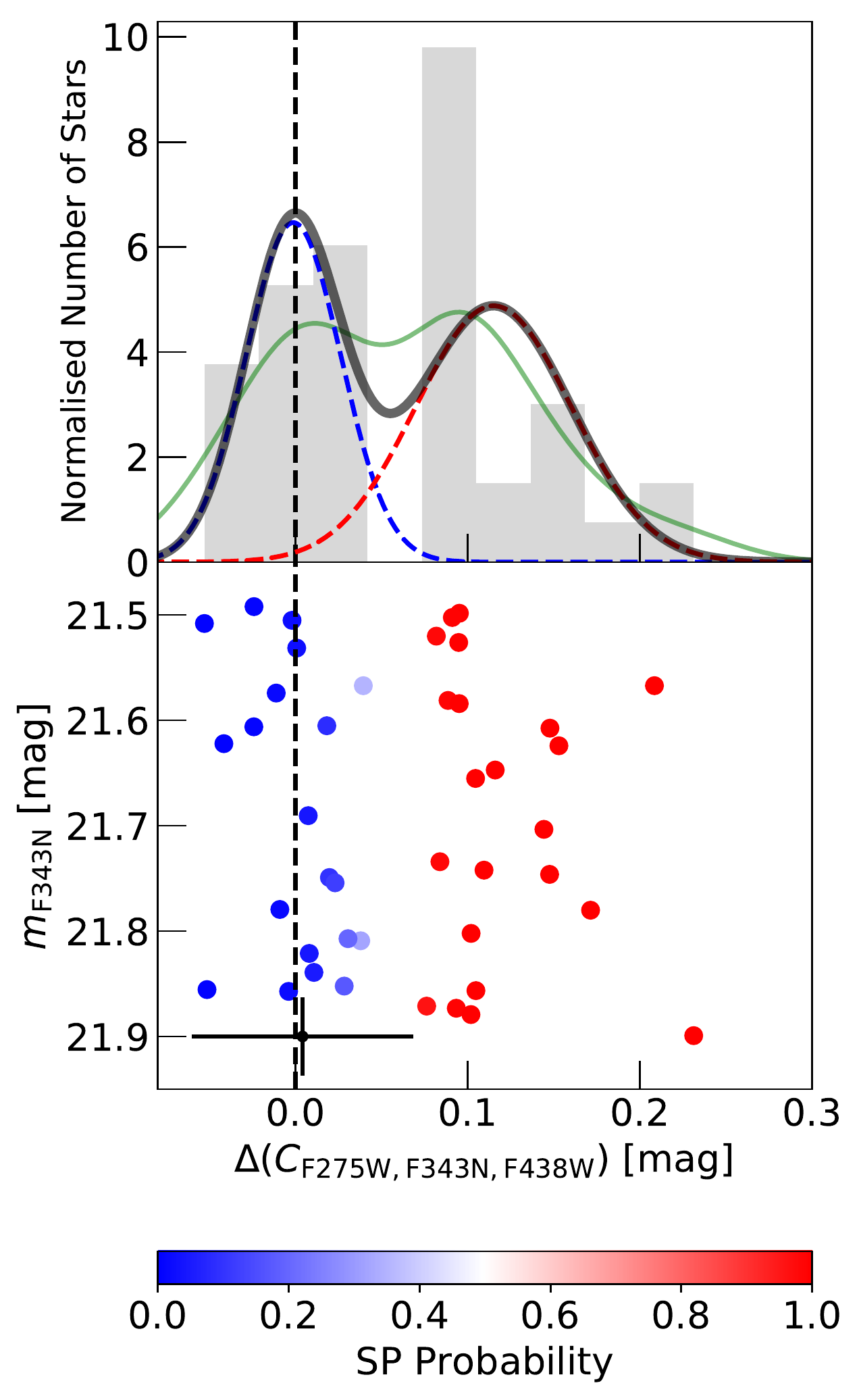}
    \caption{{\it Top Panel:} Histogram of the distribution of selected SGB stars in NGC 2121, in the $C_{F275W,F343N,F438W}$ pseudo-colour. The black solid line represents the two-component GMM best-fit function to the unbinned data. The blue (red) dashed curve represents the first (second) Gaussian component in the fit. The green curve indicates the Kernel Density Estimator (KDE). {\it Bottom Panel:} The ($m_{F343N}$, $C_{F275W,F343N,F438W}$) pseudo-colour CMD is shown, where stars are colour coded by the probability to belong to the SP. The black dashed vertical line marks the adopted fiducial line. The black errorbar shown in the bottom panel represents the typical error in $C_{F275W,F343N,F438W}$ pseudo-colours and $m_{F343N}$ magnitudes.}
    \label{fig:histo}
\end{figure}

We then calculated the distance in $C_{F275W,F343N,F438W}$ of each SGB star from the fiducial line to derive $\Delta C_{F275W,F343N,F438W}$. We fitted the rectified distribution of $\Delta C_{F275W,F343N,F438W}$ with the Gaussian Mixture Model (GMM) in order to explore the presence of two Gaussian components in the pseudo-colour distribution. The first and second Gaussian components are shown as blue and red dashed curves, respectively, in Figure \ref{fig:histo} (top panel), over the histogram of $\Delta C_{F275W,F343N,F438W}$. The solid black line instead represents the fit of the distribution using the \texttt{SCIKIT-LEARN} python package called \texttt{MIXTURE}\footnote{http://scikit-learn.org/stable/modules/mixture.html}, which applies the expectation-maximization algorithm for fitting mixture-of-Gaussian models. For comparison, we also show in green the non-parametric Kernel Density Estimate (KDE) derived from the unbinned data. It is important to stress here that what we identified as N-normal population refers to the blue dashed curve, while the N-enriched population refers to the red dashed one. We will consider hereafter the blue component as the first population (FP) in the cluster, and the red component as the second population (SP). 
We assigned to each star a probability to be part of the FP and of the SP according to the results of the GMM fit. The bottom panel of Figure \ref{fig:histo} shows the $\Delta C_{F275W,F343N,F438W}$ colours vs. $m_{F4343N}$, where the stars are colour-coded by the probability to be part of the SP. The adopted fiducial line is shown as a black dashed line.

\subsection{SGB analysis}
\begin{figure}
\centering
\includegraphics[scale=0.4]{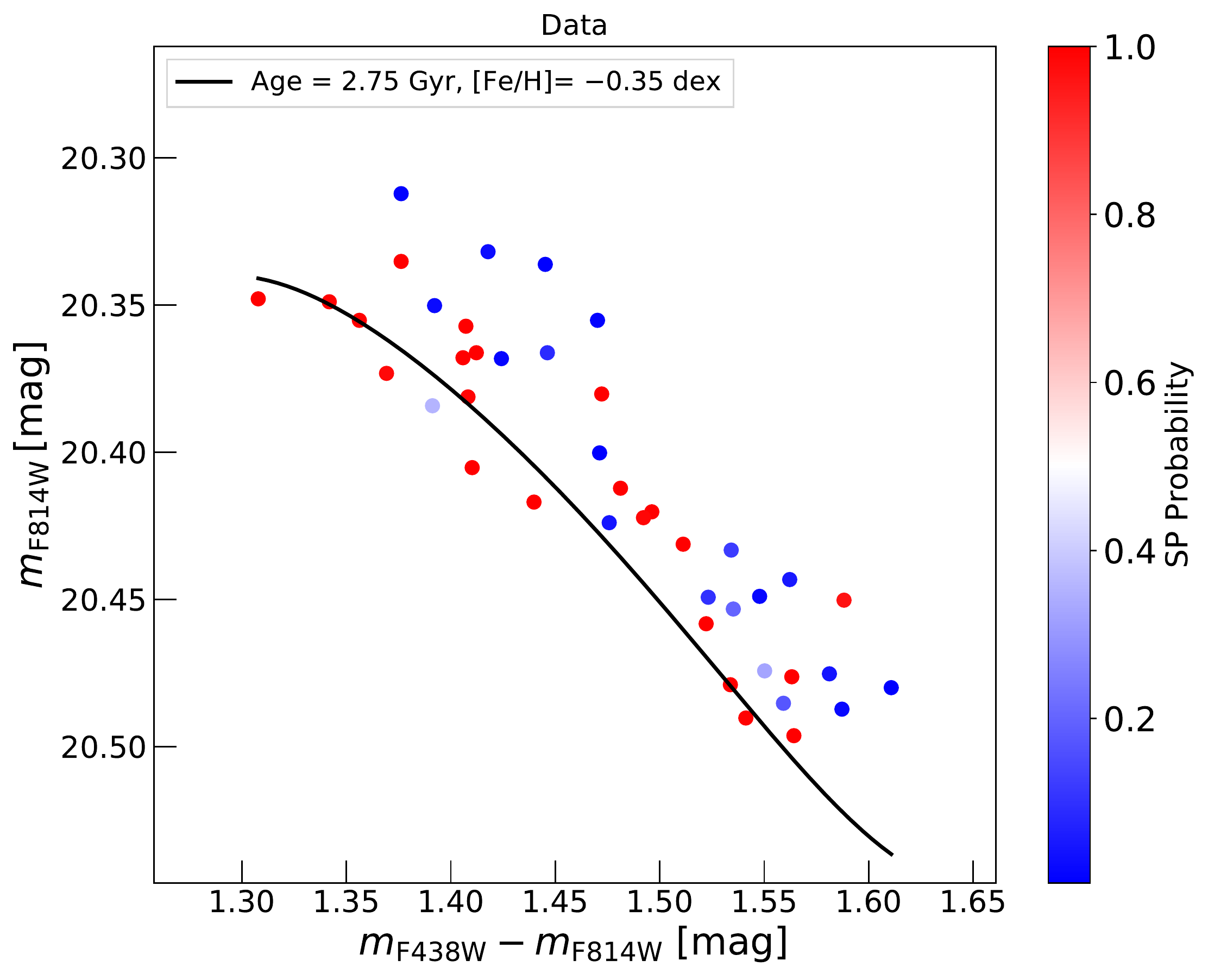}
\includegraphics[scale=0.4]{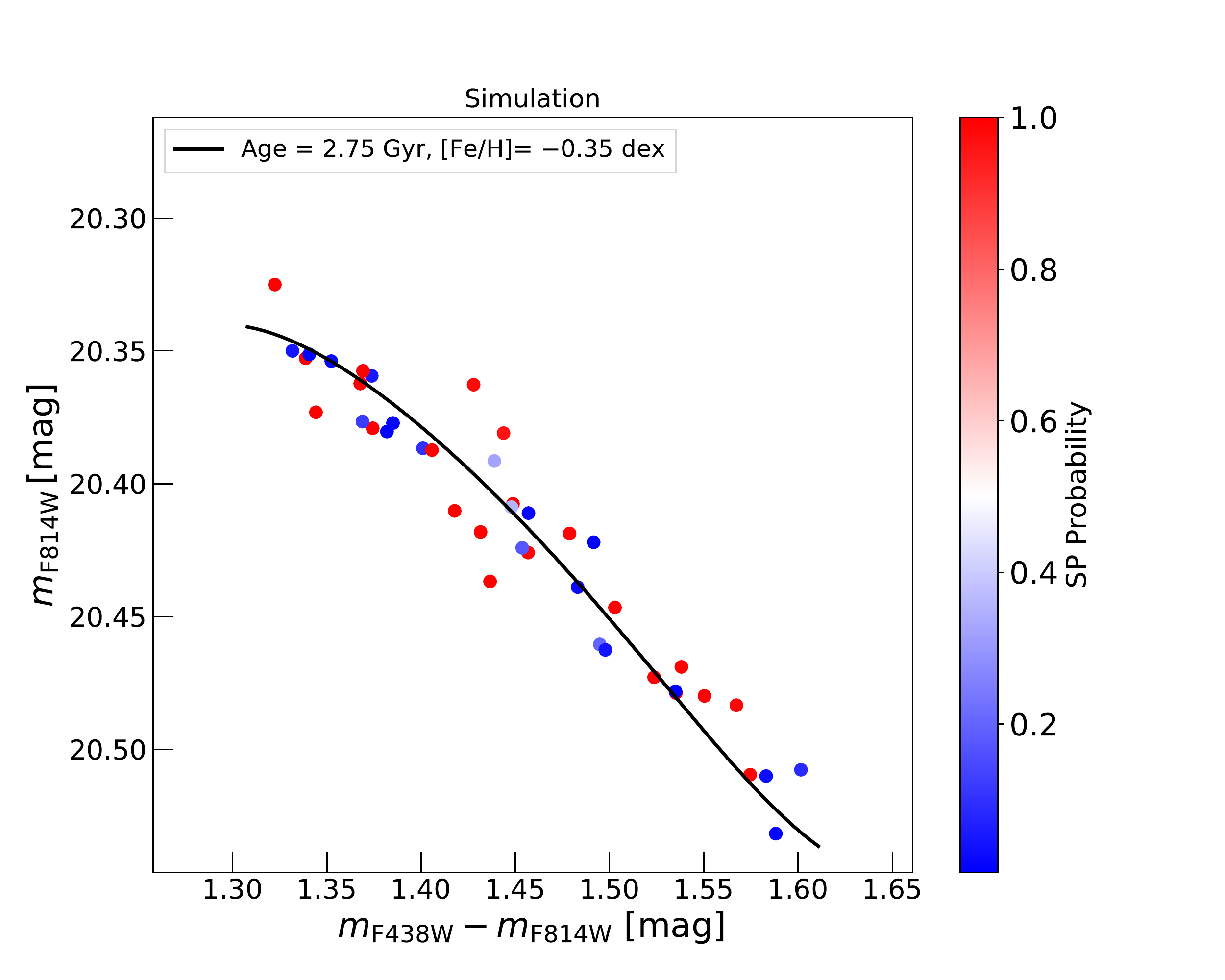}
	\caption{The final SGB stars in the ($m_{F814W}$, $m_{F438W}-m_{F814W}$) CMD for NGC 2121 are shown in the top panel, while a Monte Carlo simulation of the SGB data, where photometric errors are taken into account, are instead presented in the bottom panel. Stars are colour-coded according to the probability to be part of the SP. The black solid line indicates the defined fiducial line of the SGB based on a BaSTI isochrone having 2.75 Gyr and [Fe/H]= $-0.35$ dex.\label{fig:age_real}}
\end{figure}

In Figure \ref{fig:age_real}, top panel, the final SGB stars are shown in the optical ($m_{F814W}$, $m_{F438W}-m_{F814W}$) CMD, where all the stars are colour-coded according to their probability (from 0 to 100\%) to be part of the SP in the cluster, as derived in Section \ref{sgb1}. In this filter combination, the presence of chemical variations should not produce any significant spread or split in the SGB (e.g., \citealt{sbordone2011}). At odds, it is well known that in optical CMDs the turn-off/SGB is the ideal region to look at for possible age differences between the FP and SP, if they are present, since their SGBs should have slightly different shapes (e.g., \citealt{li2014}).

In other words, if there is an age difference between the two populations of NGC 2121, we would expect a split in the optical CMD. In this Section we explore the possible presence of such a split and then estimate the age difference through comparison with isochrones at different ages.

The top panel of Figure \ref{fig:age_real} does not show a clear offset between the populations. However, in order to address the point in a quantitative way, the idea is to compute the distance of each SGB star from a fiducial line chosen as representative of the SGB shape of the cluster, and to create from it a rectified distribution.
We defined as fiducial line the SGB part of a BaSTI isochrone \citep{pietrinferni2004}, having t = 2.75 Gyr and metallicity [Fe/H]= -0.35 dex. It is shown as a black solid curve in both panels of Figure \ref{fig:age_real}. 
The reason for using a BaSTI isochrone is the possibility to properly account for the overshooting, an effect that starts to be important in such young clusters (refer to Section 2 of \citet{martocchia2018b} for details). We adopted both distance and reddening derived from the best-fit of \citet[][]{martocchia2019}, but we had to use a slightly older age to best match our observations.
We calculated $\Delta F814W$ as the distance in the $F814W$ filter of each SGB star from the fiducial line. Then, we calculated the weighted mean in $\Delta F814W$ for FP and SP stars by using the probability to be part of one of the two populations as weights. The observed age difference between the two populations, in units of magnitudes, turns out to be $\Delta$Mag$_{OBS}$ = $\Delta F814W_{FP}$ -  $\Delta F814W_{SP}$ = (-0.024 $\pm$ 0.034) mag.

\begin{figure}
\centering
\includegraphics[scale=0.45]{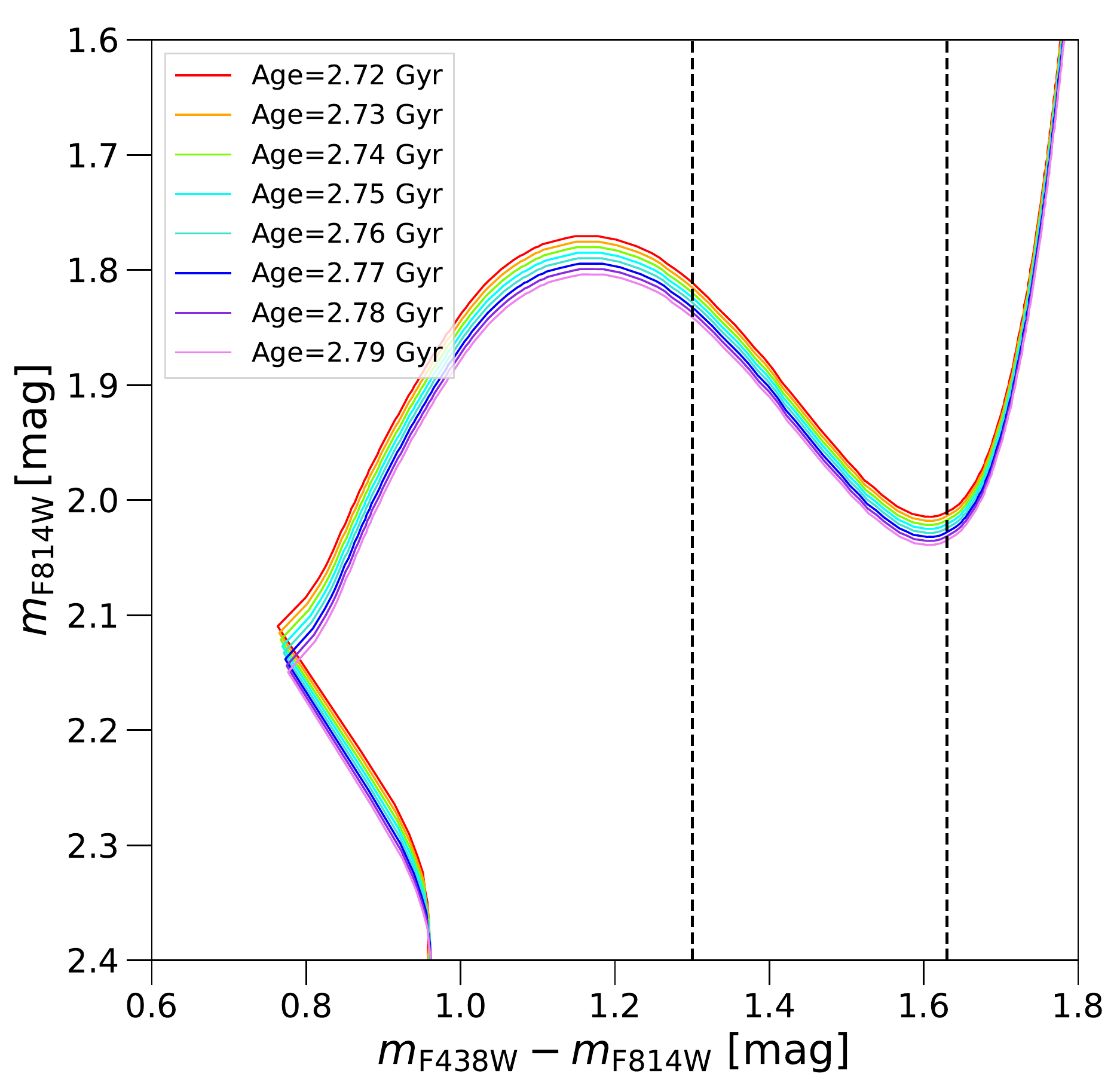}
\includegraphics[scale=0.45]{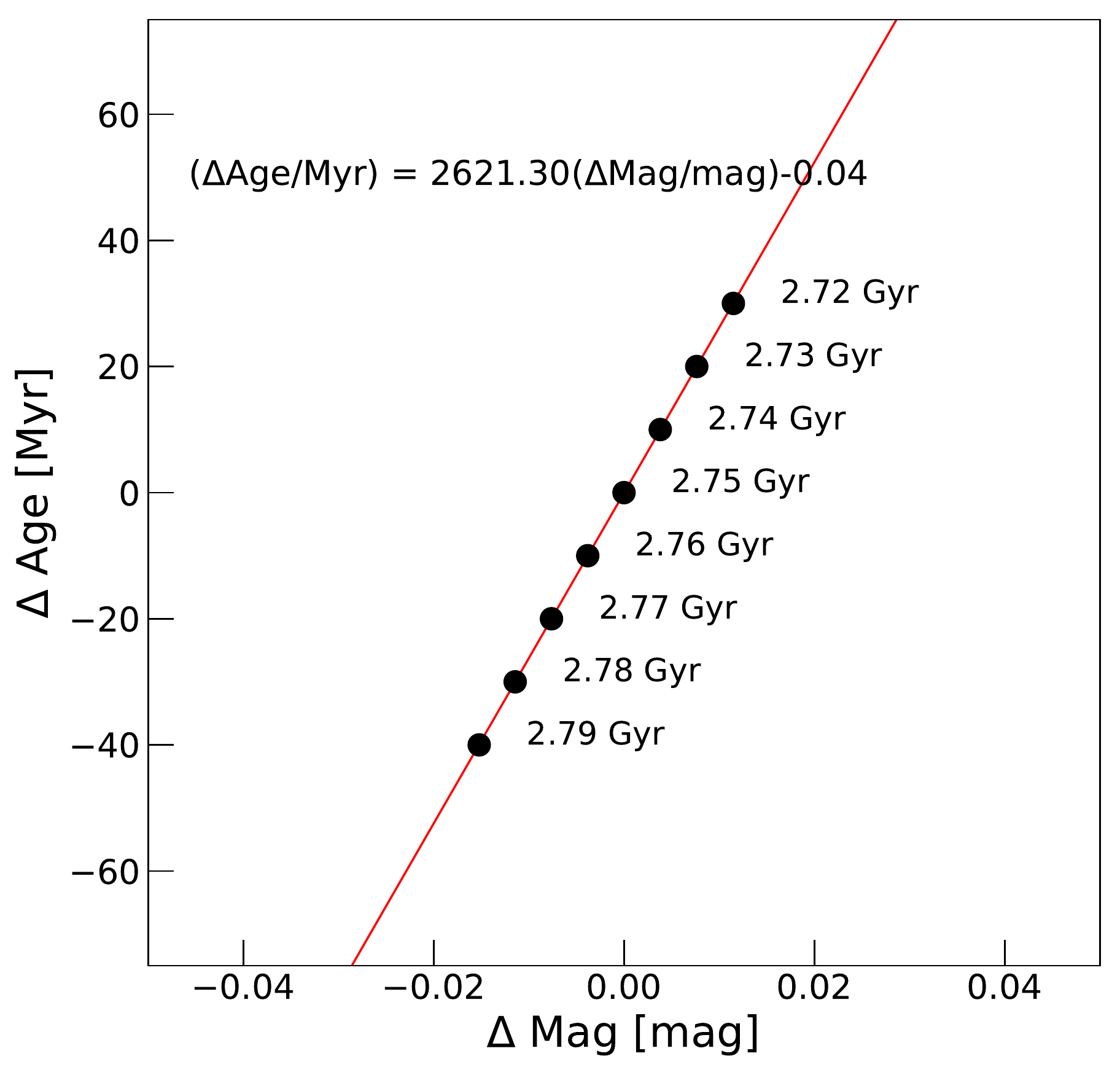}
	\caption{{\it Top panel}: BaSTI isochrones in the ($m_{F814W}$, $m_{F438W}-m_{F814W}$) CMD with ages spanning a range from 2.72 up to 2.79 Gyr, spaced by 10 Myr ([Fe/H] = $-0.35$ dex). The vertical black dashed lines define the portion of the SGB used to compute the difference in age between the two populations. {\it Bottom panel}: The relation between $\Delta$ Mag and $\Delta$ Age for the SGB is presented as a solid red. The black circles indicate the values obtained for the isochrones shown in the top panel. See the text for more details on the procedure.\label{fig:ageiso}}
\end{figure}

This difference in magnitude between the two populations can be easily converted in age difference in units of Myr by comparing this result with appropriate evolutionary models. As already discussed, we used BaSTI isochrones, that we interpolated from age = 2.72 Gyr to 2.79 Gyr, spaced by 10 Myr. The adopted isochrones are shown in the top panel of Figure \ref{fig:ageiso}, where two vertical dashed lines indicate the selected region of the SGB used to perform the analysis. This region is exactly the one where a bimodal distribution in the pseudo-colour diagram has been identified. We first used the 2.75 Gyr isochrone as an age reference for NGC 2121. Then we computed the mean difference in $F814W$ magnitudes, at the SGB level, between the reference isochrone and the others (shown in Figure \ref{fig:ageiso}). From these, presented as black circles in the bottom panel of Figure \ref{fig:ageiso}, we derived a relation through linear fitting (red solid line) which can be used to transform the observed difference in magnitude to age difference ($\Delta$Age) in Myr:
\begin{equation}
\Delta Age [Myr] = 2621.30 \times \Delta Mag [mag] - 0.04
\end{equation}
By adopting this relation we found that the age difference in the ($m_{F814W}$, $m_{F438W}-m_{F814W}$) CMD between the first and the second population of NGC 2121 is -6 Myr, with the SP slightly younger than the FP.

To estimate the uncertainty of this result, taking into account the limited number of available stars, we made used of Monte Carlo simulations. Starting from a single isochrone model, we simulated a sample having the same number of stars as the ones previously adopted, we added their photometric errors and we assigned to each of these stars the same probability distribution of belonging to the FP and the SP as in the real data. We repeated this process 100,000 times to be statistically robust. 
One of these simulations is reported, as an example, in the bottom panel of Figure \ref{fig:age_real} and as can be seen, it reproduces pretty well the observations. For each simulation, we calculated the simulated $\Delta F814W$ for each population and thus $\Delta$Mag$_{OBS}$. We found that the Monte Carlo distribution has a mean of 0 mag, as expected, and $\sigma$ = 0.0046 mag, which translates into an age difference of 12 Myr, computed using Equation 1.

We found that the age difference between the two populations in the SGB of NGC 2121 is then -6 $\pm$ 12 Myr, which is consistent with no age spread. In order to be sure that an error in the determination of the best-fit age of the cluster does not affect our conclusions, we took ages of 2.0 Gyr and 3.0 Gyr as reference to do the same exercise. In the first case we found an age difference between FP and SP of -4.6 $\pm$ 6 Myr, while in the second one the age difference turned out to be of -8 $\pm$ 14 Myr. Both results confirm that the observed difference in age between the two populations of NGC 2121 is small and consistent with zero, regardless the isochrone used to make such an estimate. 

It is worth to mention that star-to-star variations of He abundance might also generate an offset on the SGB \citep{kim2018}. At these ages, in particular, the SGB of an He-enriched population appears slightly fainter than the SGB of a He-normal population, thus contributing to compensate in some way the effect of an age difference between the two. In order to quantify the impact of an He spread on the SGB analysis, we compared two isochrones from the Victoria-Regina stellar models \citep{casagrande2014}, having age and metallicity of NGC 2121 but a slightly different He content (Y=0.256 for the He-normal population and Y = 0.276 for the He-enriched one), following what have been derived in Section \ref{sec:helium}. We found that the isochrones are indistinguishable in the SGB portion (1.3 $< m_{F438W}-m_{F814W} <$ 1.65; see Figure \ref{fig:ageiso}, top panel) we used to compute the age differences between the two populations, so that the presence of a possible He spread does not have any impact on our estimate of $\Delta$Age$_{P1-P2}$.

\section{Discussion and Conclusions}
\label{sec:concl} 
In this work we made use of high-resolution near-UV/optical archival and proprietary HST observations to study in detail the presence of MPs in the LMC clusters NGC 2121 and NGC 1783, having ages of $\sim$2.5 Gyr and $\sim$1.5 Gyr, respectively. 
In particular we exploited the power of the F343N filter to create an alternative ``chromosome map" \citep{milone2017}, which is able to efficiently separate sub-populations of stars having different N-abundances. By comparing these chromosome maps with that of the intermediate-age ($\sim$7.5 Gyr) cluster Lindsay 1 \citep{saracino2019a}, we found a clear correlation between the width ($\sigma$) of the $\Delta_{F275W,F343N,F438W}$ distribution and the age of the cluster itself, with larger $\sigma$ associated to older ages. This finding well agrees with the results found by \citet[and references therein]{martocchia2019} on a large sample of LMC/SMC clusters. However, this no longer implies that the N-enhancement within a cluster decreases going from old to young ages because of the effect of the first dredge-up which modifies the observed $\Delta_{F275W,F343N,F438W}$ in a not straightforward way \citep{Salaris2020}.

The construction of chromosome maps for clusters belonging to (non-accreted) galaxies outside the MW has been recently possible (\citealt{saracino2019a,sills2019,milone2019} show few examples). This is a very good opportunity since the one-to-one comparison allows to show up possible similarities and/or differences among clusters belonging to galaxies having different morphologies and star formation histories. In this respect, what already found by \citet[][]{saracino2019a} and here confirmed for NGC 2121 and NGC 1783 is that, in the chromosome maps of LMC/SMC clusters, the N-normal and N-enriched populations appear to follow a rather continuous sequence, instead of creating two well separated clumps, as many MW GCs do. At odds, all of them occupy the same parameter space, with almost the same shape as MW GCs. This finding, if further observed in other LMC/SMC clusters, could open to new and still unexplored questions. 

Since these clusters have different metallicities, it would be then interesting to compare them in a metallicity-free parameter space. However the metallicity-dependent corrections used by \citet[][]{marino2019} to create a ``universal'' chromosome map and tested by \citet{sills2019} on the Sagittarius dwarf cluster M54, only work for clusters where the position of the first population can be easily determined. It is not the case for LMC/SMC clusters.

Finally, we focused a bit more in detail on NGC 2121, by exploring two main features: {\it 1)} the RC morphology, in order to infer the He enrichment of the cluster; and {\it 2)} the structure of the SGB, in order to investigate whether the N-normal and N-enriched populations show different ages.

The RC study of NGC 2121 revealed that the effects of a He enrichment and a differential mass-loss along the RGB are somehow coupled for clusters having such an age and metallicity. Indeed, two possible scenarios can most likely take place in such a cluster: \\{\it 1)} $\Delta$Y$_{ini}$=0.0 and $\Delta M_{RGB}$=0.03-0.04;\\ {\it 2)} $\Delta$Y$_{ini}$=0.020$\pm$0.005 and $\Delta M_{RGB}$=0.0. 

We cannot give a firm answer to which one should be preferred since they are statistically indistinguishable, but we can consider these values as upper limits since a combination of these two effects could also be in place in NGC 2121.

From the SGB analysis we found that $C_{F275W,F343N,F438W}$ is a powerful pseudo-colour to identify FP and SP even at the SGB level, thus allowing to easily distinguish between them. By using the optical ($m_{F814W}$, $m_{F438W}-m_{F814W}$) CMD we then derived an age difference of -6 $\pm$ 12 Myr, which is relatively small (with the SP slightly younger than the FP), and consistent with zero within the uncertainties. This result turns out to be in very good agreement with what found in the $\sim$2 Gyr old LMC star cluster NGC 1978 by \citet[][]{martocchia2018b}, the only other cluster where such a test has been possible, thus strengthening the constraints on the onset of MPs and on the source of the chemical enrichment within clusters. Indeed, theories predicting an age difference from at least 30 Myr up to 200 Myr between the first and second populations appear to be ruled out from such experiments. Further theoretical efforts seem to be required in order to explain all the recent findings in the context of a general framework.

\section*{Acknowledgements}
We acknowledge the anonymous referee for the detailed review and helpful suggestions, which allowed us to improve the manuscript. SS and NB gratefully acknowledge financial support from the European Research Council (ERC-CoG-646928, Multi-Pop). NB also acknowledges support from the Royal Society (University Research Fellowship). WC acknowledges funding from the Swiss National Science Foundation under grant P2GEP2 171971. V.K-P. is very gratitude to Jay Anderson for sharing with us his ePSF code. CL thanks the Swiss National Science Foundation for supporting this research through the Ambizione grant number PZ00P2 168065. The authors gratefully acknowledge financial support for programs GO-14069, GO-15062 and GO-15630, provided by NASA through Hubble Fellowship grant HST-HF2-51387.001-A awarded by the Space Telescope Science Institute, which is operated by the Association of Universities for Research in Astronomy, Inc., for NASA, under contract NAS5-26555.



\bibliographystyle{mnras}
\bibliography{lind1}

\bsp	
\label{lastpage}
\end{document}